# The role of surface charge in the interaction of nanoparticles with model pulmonary surfactants

F. Mousseau[*] and J.-F. Berret[*]

*Matière et Systèmes Complexes, UMR 7057 CNRS Université Denis Diderot Paris-VII, Bâtiment Condorcet, 10 rue Alice Domon et Léonie Duquet, 75205 Paris, France.*

**Abstract:** Inhaled nanoparticles traveling through the airways are able to reach the respiratory zone of the lungs. In such event, the incoming particles first enter in contact with the liquid lining the alveolar epithelium, the pulmonary surfactant. The pulmonary surfactant is composed of lipids and proteins that are assembled into large vesicular structures. The question of the nature of the biophysicochemical interaction with the pulmonary surfactant is central to understand how the nanoparticles can cross the air-blood barrier. Here we explore the phase behavior of sub-100 nm particles and surfactant substitutes in controlled conditions. Three types of surfactant mimetics, including the exogenous substitute Curosurf®, a drug administred to infants with respiratory distress syndrome are tested together with aluminum oxide ($Al_2O_3$), silicon dioxide ($SiO_2$) and polymer (latex) nanoparticles. The main result here is the observation of the spontaneous nanoparticle-vesicle aggregation induced by Coulombic attraction. The role of the surface charges is clearly established. We also evaluate the supported lipid bilayer formation recently predicted and find that in the cases studied these structures do not occur. Pertaining to the aggregate internal structure, fluorescence microscopy ascertains that the vesicles and particles are intermixed at the nano- to microscale. With particles acting as stickers between vesicles, it is anticipated that the presence of inhaled nanomaterials in the alveolar spaces could significantly modify the interfacial and bulk properties of the pulmonary surfactant and interfere with the lung physiology.

**Keywords**: Nanoparticles – Pulmonary surfactant – Curosurf[®] – Bio-nano interfaces – Electrostatic interactions – Supported Lipid Bilayers

# I – Introduction

Recent studies have shown the existence of an association between the exposure to air pollution and the increase of mortality from cardiorespiratory diseases.[1] Lung associated diseases have in general multifactorial backgrounds, mostly depending on the interaction of environmental and genetic factors.[2] The deposition of inhaled particles in the respiratory track has been identified as one of these factors. It is now well established that particles with sizes below 100 nm are able to reach the respiratory zone in significant amounts and cross the air-blood barrier, entering then into the systemic circulation.[3,4] In the alveolar spaces, incoming particles come in contact with a thin (< 1 μm) layer of the fluid lining the alveolar epithelium, the pulmonary surfactant. The pulmonary surfactant is essential for the lung physiology as it reduces the surface tension with the alveolar cells and represents the first barrier against pathogens. The pulmonary surfactant is a complex surface-active fluid that contains phospholipids and proteins in a ratio 90:10 at a total concentration of 35 g L$^{-1}$.[5] At the air-liquid interface, the lipids are assembled into a monolayer whereas beneath the surface (in the hypophase) they form vesicle-like aggregates, such as multivesicular vesicles, lamellar bodies and tubular myelin.[5-7]

To evaluate the risks of exposure to inhaled nanomaterials, recent experimental, theoretical and simulations studies have focused on the interaction of nanoparticles with unilamellar vesicles.[8-23] Depending on the particle size, several mechanisms have been proposed, leading to a wide variety of predictions including particles decorating the outer vesicle surface, supported lipid bilayers (SLB) or particles more generally engulfed in vesicles.[16,19,20,23,24] In the latter case, the fluid membrane invaginates and envelops one or several particles like in cellular endocytosis.[16,21,22] SLBs on highly curved surfaces have been the focus of many recent studies, as they represent a means to coat particles with biocompatible interfaces. Drug delivery platforms based on porous particles and decorated with a lipid membrane have been developed from different groups, achieving hybrid structures with remarkable stability and cellular uptake properties.[10,25-27] With hydrosoluble particle, the wrapping of bio-membranes is favored if the adhesive interactions are sufficiently strong to compensate the bending energy.[16,21,22,28-31] In contrast, with hydrophobic nanoparticles the formation of a single monolayer corona or the insertion into the membrane is predicted.[16,29]

The great majority of the experimental work done so far was performed using synthetic phospholipids (e.g. DPPC,[31] DMPC[13] or DOPC,[9,14,15,23]) or mixtures thereof.[10,12,25,32-35] Here DPPC, DMPC and DOPC stand for 1,2-dipalmitoylphosphatidylcholine, 1,2-dimyristoyl-*sn*-glycero-3-phosphocholine and 1,2-dioleoyl-*sn*-glycero-3-phosphocholine respectively. Although the vesicular structures achieved with synthetic lipids are similar to those found in endogenous surfactant, they lack the lipid variety present in natural surfactant as well as the membrane proteins SP-A, SP-B, SP-C and SP-D. A different approach consists in using existing exogenous surfactants such as those administered to newborns with respiratory distress syndrome.[26,27,34,36-43] These surfactant formulations are of porcine (Curosurf®, Survanta®) or bovine (Alveofact®) origin and possess both SP-B and SP-C hydrophobic proteins. From a practical perspective, substitutes are easier to handle compared to the endogenous surfactant and are considered as reliable surfactant models. In particular they exhibit excellent temporal stability, allowing to perform extended characterization and interaction experiments.[34] Despite many efforts, the mechanisms of particles interacting with the pulmonary surfactant are not fully understood. The question of the nature of the physico-chemical interactions is central, as these interactions modify the particle biological identity and regulate its transfer through the air-blood barrier.

In the present paper, we provide an extended physico-chemical study of the interaction between Curosurf® vesicles and 50 nm synthetic nanoparticles made from aluminum oxide (Al$_2$O$_3$), silicon dioxide (SiO$_2$) or polymers (latex). To this aim, we use the method of continuous variation developed by Paul Job to determine the stoichiometry of binding (macro)molecular species in solutions.[44,45] The method is here combined with static and dynamic light scattering, leading to Job scattering plots.[34,46-49] Emphasis is put on the formation of nanoparticle-vesicle aggregates and of supported lipid bilayers, which have both attracted attention in the context of particles crossing the air-blood barrier in the lungs.[10,20,24,50] For the first time, we provide predictions for the SLB formation in the Job representation and compare them with experimental results. The conclusion that emerges from the survey is that nanoparticles and lipid vesicles interact predominantly *via* electrostatic charge mediated attraction and do not form supported lipid bilayers spontaneously. Co-localization fluorescence and cryogenic transmission electron microscopy on 42 nm aminated silica confirm these conclusions, although these findings also evidence a broad variety of novel structures at the nanoscale.

## II – Experimental

### II.1 – Materials

*Nanoparticles*: Aluminum oxide nanoparticle powder from Disperal® (SASOL, Germany) was dissolved in a nitric acid solution (0.4 wt. % in deionized water) at the concentration of 10 g L$^{-1}$ and sonicated for an hour. The dispersion was diluted down to 0.1 g L$^{-1}$ and the pH adjusted at 5.[34] The nanoalumina particle have the shape of irregular platelets of sizes 40 nm in length and 10 nm in thickness and their hydrodynamic diameter was measured at $D_H$ = 64 nm. The positively charged silica particles were synthetized using the Stöber synthesis.[51] Briefly, fluorescent silica seeds were prepared in three steps. The rhodamine derivative, rhodamine red c2 maleimide (Aldrich) was first covalently bound to the silica precursor (3-mercaptopropyl)trimethoxysilane (MPS, Aldrich). The rhodamine-MPS compound was then mixed with tetraethyl orthosilicate silica precursor (TEOS, Aldrich) and the Stöber synthesis was performed. With this approach the dyes were covalently bound to the silica matrix. In a third step, a non-fluorescent silica shell was grown with TEOS to increase the particle size and prevent the leaking of the dyes out of the particles. Functionalization by amine groups was then performed, resulting in a positively charged coating.[51] Aminated silica were synthesized at 40 g L$^{-1}$ and diluted with DI-water at pH 5. The hydrodynamic and geometric diameters were determined at $D_H$ = 60 nm and $D_0$ = 41.2 nm. Fluorescence properties were characterized using a Cary Eclipse fluorimeter (Agilent), leading excitation and emission peaks at 572 nm and 590 nm respectively. Negative silica particles (trade name CLX®) were purchased from Sigma Aldrich at the concentration of 450 g L$^{-1}$. The batch was diluted down to 50 g L$^{-1}$ and dialyzed against DI-water at pH 9 for two days. The diameters were measured at $D_H$ = 34 nm and $D_0$ = 20 nm. Latex particles functionalized with carboxylate or amidine groups were obtained from Molecular Probes (Invitrogen, USA). The initial 40 g L$^{-1}$ batches were diluted keeping the pH of the anionic and cationic dispersions constant at pH 9 and pH 6. The hydrodynamic (resp. geometric) diameters were $D_H$ = 39 nm (resp. $D_0$ = 30 nm) for the carboxylate modified particles and $D_H$ = 56 nm (resp. $D_0$ = 34 nm) for the amidine functionalized particles. The differences found between the geometric and hydrodynamic diameters are attributed to the particle size dispersity. For sake of simplification, the previous particles will be abbreviated Alumina (+), Silica (+), Latex (+), Latex (-) and Silica (-) throughout the paper. The surface charge density for the 5 particles was finally determined using the polyelectrolyte assisted charge titration spectrometry at $\sigma$ = +7.3$e$, +0.62$e$, +0.33$e$, -0.048$e$ and -0.31$e$ nm$^{-2}$, respectively.[48] Table I summarizes the bulk and surface properties.

| Nano-particle | Chemical composition | Function-nalization | $D_0$ (nm) | $s$ | $D_H$ (nm) | $\sigma$ (nm$^{-2}$) |
|---|---|---|---|---|---|---|
| Alumina (+) | Al$_2$O$_3$ | / | 40 | 0.3 | 64 | +7.3$e$ |
| Silica (+) | SiO$_2$ | Amine | 42 | 0.11 | 60 | +0.62$e$ |
| Latex (+) | Polystyrene | Amidine | 34 | 0.15 | 56 | +0.33$e$ |
| Latex (-) | Polystyrene | Carboxylate | 30 | 0.15 | 39 | -0.048$e$ |
| Silica (-) | SiO$_2$ | / | 20 | 0.20 | 34 | -0.31$e$ |

**Table I**: List of nanoparticles and their characteristics. $D_0$ and $D_H$ stand for the geometric and hydrodynamic diameters determined by transmission electron microscopy and dynamic light scattering. $s$ denotes the size dispersity (ratio between the standard deviation and average size of the distribution). The electrostatic charge densities $\sigma$ was obtained using polyelectrolyte assisted charge titration spectrometry.[48]

*Exogeneous pulmonary surfactant*: Curosurf®, also called Poractant Alfa (*Chiesi Pharmaceuticals*, Parma, Italy) is an extract of whole mince of porcine lung tissue purified by column chromatography.[36] It is produced as a 80 g L$^{-1}$ phospholipid and protein suspension containing, among others phosphatidylcholine lipids, sphingomyelin, phosphatidylglycerol and the membrane proteins SP-B and SP-C.[52] The Curosurf® lipid and protein compositions are provided in **Supplementary Information S1** and compared to that of native surfactant obtained by saline bronchoalveolar lavage.[37,38,52] Curosurf® is indicated for the rescue treatment of Respiratory Distress Syndrome (RDS) in premature infants and is administered at a dose of 200 mg per kilogram. According to the manufacturer, the pH of Curosurf® is adjusted with sodium bicarbonate at pH 6.2.[34,37] Curosurf® was kindly provided by Dr. Mostafa Mokhtari and his team from the neonatal service at Hospital Kremlin-Bicêtre, Val-de-Marne, France.

*Synthetic pulmonary surfactant*: Dipalmitoylphosphatidylcholine (DPPC), L-α-Phosphatidyl-DL-glycerol sodium salt from egg yolk lecithin (PG, Sigma-Aldrich, MDL number: MFCD00213550) and 2-Oleoyl-1-palmitoyl-sn-glycero-3-phospho-rac-(1-glycerol) (POPG) were purchased from Sigma-Aldrich. Phospholipids DPPC, PG and POPG were initially dissolved in methanol, at 10, 10 and 20 g L$^{-1}$ respectively and then mixed in proper amounts for a final weight concentration of 80% / 10% / 10% of DPPC / PG / POPG. The solvent was evaporated under low pressure at 60 ˚C for 30 minutes. The lipid film formed on the bottom of the flask was then rehydrated with the addition of Milli-Q water at 60 ˚C and agitated at atmospheric pressure for another 30 minutes. Milli-Q water was added again to finally obtain a solution at 1 g L$^{-1}$. For the DPPC/PG/POPG mixture, it is assumed that the gel-to-fluid transition is closed to that of DPPC ($T_M$ = 41 °C).[10]

*Membrane labeling*: PKH67 (Sigma Aldrich) is a molecule composed of a green fluorescent head and an aliphatic chain. It is commonly used to stain living cells by incorporation into the membranes.[53] This dye is used here to label surfactant vesicles without drying or use of an organic solvent. For the labeling vesicles, the dye concentration was adjusted to prevent alteration of the membrane properties. Curosurf (resp. PKH67) was diluted with DI-water at 2 g L$^{-1}$ (resp. 2×10$^{-6}$ M) and mixed rapidly in equal volumes. In such conditions, the dye/lipid ratio was 1:1400, indicating a large excess of Curosurf® molecules. The dispersion was further kept 15 min in the dark to ensure complete dye insertion in the membranes. Fluorescence properties were characterized using a Cary Eclipse fluorimeter (Agilent), leading excitation and emission peaks at 489 nm and 504 nm respectively. Interaction strength measurements using scattering Job plots were performed and similar interaction strengths for vesicles with and without dye molecules were found. The overall results show that PKH67 dyes do not interfere with the vesicular structure and stability (**Supplementary Information S2**).

## II.2 – Experimental techniques

*Mixing*: The interaction between pulmonary surfactants and nanoparticles was investigated using a mixing protocol known as the continuous variation method.[44,46,48,54] Surfactant and particle batches (10 mL) were prepared in the same conditions of pH and concentration (0.1 or 1 g L$^{-1}$). The solutions were mixed at different volumetric ratios $X = V_{Surf}/V_{NP}$ where $V_{Surf}$ and $V_{NP}$ are the volumes of the surfactant and particle solutions respectively. The stock concentrations being alike, the volumetric ratio $X$ is also equivalent to the mass or to the concentration ratios. The continuous variation method allows exploring a broad range of mixing conditions while keeping the total concentration in the dilute regime.[46,55] Requirements on the pH were imposed by the fact that the conditions encountered in alveolar spaces are mildly acidic (pH 6.4),[5] and by the particle and vesicle physico-chemical properties. For Alumina (+) and Latex (-), the pH for the stock dispersions was adjusted to ensure that particles do not aggregate as a result of the pH changes. Silica (+), Latex (+) and Silica (-) were studied at physiological pH, Alumina (+) at pH 5 and Latex (-) at pH 9. Table II lists the pHs at the working conditions tested for the different surfactants and shows that the zeta potentials are negative, from $\zeta$ = - 23 mV to $\zeta$ = - 61 mV.

*Extrusion*: Curosurf® extrusion was performed using an Avanti Mini Extruder (Avanti Polar Lipids, Inc. Alabama, USA). Solutions were prepared at 1 g L$^{-1}$ and extruded 50 times through Whatman Nucleopor polycarbonate membranes of different pore sizes: 50, 100, 200, 400 and 800 nm. The extrusion device and membrane scission mechanism, as well as the results of the vesicle diameter *versus* pore sizes are summarized in **Supplementary Information S3**.

| Pulmonary surfactant | Working pH | Zeta potential (mV) |
|---|---|---|
| Curosurf® | 5 | - 38 ± 5 |
| | 6.4 | - 54 ± 8 |
| | 9 | - 61 ± 6 |
| Extruded Curosurf® | 5 | - 23 ± 10 |
| | 6.4 | - 23 ± 10 |
| | 9 | - 32 ± 11 |
| Synthetic | 6.4 | - 36 ± 12 |

**Table II**: Zeta potentials of the surfactant vesicles in the pH conditions tested. Silica (+), Latex (+) and Silica (-) were studied at the alveolus physiological pH (pH 6.4), Alumina (+) at pH 5 and Latex (-) at pH 9.

***Static and Dynamic Light Scattering***: The scattered intensity $I_S$ and the hydrodynamic diameter $D_H$ were obtained from the NanoZS Zetasizer spectrometer (Malvern Instruments). The second-order autocorrelation function was analyzed using the cumulant and CONTIN algorithms to determine the average diffusion coefficient $D_C$ of the scatterers. $D_H$ was calculated according to the Stokes-Einstein relation $D_H = k_B T / 3\pi\eta D_C$ where $k_B$ is the Boltzmann constant, $T$ the temperature and $\eta$ the solvent viscosity. Except otherwise mentioned, the hydrodynamic diameters provided here are the second coefficients in the cumulant analysis described as $Z_{Ave}$. Measurements were performed in triplicate at 25 °C and 37 °C after an equilibration time of 120 s, yielding experimental uncertainties better than 10% in both intensity and diameter.

***Electrophoretic mobility and zeta potential***: Laser Doppler velocimetry using the phase analysis light scattering mode and detection at an angle of 16° was used to carry out the electrokinetic measurements of electrophoretic mobility and zeta potential with the Zetasizer Nano ZS equipment (Malvern Instruments, UK). Zeta potential was measured after a 120 s equilibration at 25 °C.

***Transmission electron microscopy***: TEM imaging was performed with a Tecnai 12 operating at 80 kV equipped with a 1k×1k Keen View camera. 3 µL of a nanoparticle-surfactant dispersion (concentration 0.05 g L$^{-1}$) were deposited on holey-carbon coated 300 mesh copper grids (Neyco). The sample was stained during 30 s with uranyl acetate at 0.5 wt. %. Uranyl acetate provides a dark labeling of the phospholipid heads due to the high affinity of the electron dense ion uranyl to carboxyl groups. Grids were let to dry at room temperature in the dark for 20 min. For cryogenic transmission electron microscopy (cryo-TEM), few microliters of surfactant (concentration 5 g L$^{-1}$ in DI-water) or silica–surfactant dispersions (concentration 0.5 g L$^{-1}$ in white DMEM) were deposited on a lacey carbon coated 200 mesh (Ted Pella Inc.). The drop was blotted with a filter paper using a FEI Vitrobot$^{TM}$ freeze plunger. The grid was then quenched rapidly in liquid ethane to avoid crystallization and later cooled with liquid nitrogen. It was then transferred into the vacuum column of a JEOL 1400 TEM microscope (120 kV) where it was maintained at liquid nitrogen temperature thanks to a cryo-holder (Gatan). The magnification was comprised between 3000× and 40000×, and images were recorded with an 2k×2k Ultrascan camera (Gatan). TEM and cryo-TEM images were digitized and treated by the ImageJ software and plugins (http://rsbweb.nih.gov/ij/).

*Optical microscopy*: Images were acquired on an IX73 inverted microscope (Olympus) equipped with an 60× objective. An EXi Blue camera (QImaging) and Metaview software (Universal Imaging Inc.) were used as the acquisition system. The illumination system "Illuminateur XCite Microscope" produced a white light, filtered for observing a red (excitation filter at 545 nm - bandwidth 30 nm and emission filter at 620 nm - bandwidth 60 nm) or a green (excitation filter at 470 nm - bandwidth 40 nm and emission filter at 525 nm - bandwidth 50 nm) signal in fluorescence. Fluorescent positive silica particles and labelled Curosurf® were diluted in DI-water to 1 g L$^{-1}$. Solutions pre-heated at 37 °C were mixed at $X = 0.4$ and kept over night at physiological temperature. Thirty microliters of dispersion (diluted 20 times with DI-water at 37 °C) were deposited on a glass plate (previously equilibrated at physiological temperature for 15 min) and sealed into a Gene Frame dual adhesive system (Abgene/ Advanced Biotech). During observation, the slide was placed in a chamber (Heating Insert P Lab-Tek$^{TM}$, PECON) thermostated at 37°C by the Tempcontrol 37-2 digital module (PECON).

## III – Results and discussion
### III.1 – Nanoparticles and pulmonary surfactants

Figs. 1a-e display transmission electron microscopy images of the different nanoparticles. Alumina have the shape of irregular platelets of dimensions 40 nm in length and 10 nm in thickness, whereas silica and latexes are spherical and characterized by diameters of 20 - 40 nm and dispersities are in the range 0.1 - 0.2 (Table I). In solutions, the colloidal stability is ensured by electrostatic repulsion arising from point surface charges and electrostatic double layer.[56] The particles were selected to cover a broad range of charge densities noted $\sigma$, from +7.3$e$ nm$^{-2}$ for Alumina (+) to -0.31$e$ nm$^{-2}$ for Silica (-). The $\sigma$'s were determined experimentally using the polyelectrolyte assisted charge titration spectrometry method published recently.[48]

Fig. 1f displays a representative cryo-transmission electron microscopy image of Curosurf® at 5 g L$^{-1}$. Additional images are available in **Supplementary Information S4**. The figure shows that the phospholipids self-assemble into unilamellar and multivesicular vesicles.[38,43] Multivesicular vesicles describe large membrane compartments encapsulating one or several smaller vesicles. At the scale of the membrane, the lipids form a bilayer structure of thickness $\delta$ = 4.4 nm (**Supplementary Information S5**). The vesicle size distribution derived from cryo-TEM (taking into account all enclosed structures) extends from 50 to 1000 nm, with a median value of 230 nm and a dispersity of 0.55. The vesicular structures observed for Curosurf® are similar to those reported in the literature for endo- and exogenous pulmonary surfactant.[27,37,52,57]

Fig. 1g provides examples of Curosurf® vesicles obtained from a 1 g L$^{-1}$ extruded dispersion, using here a 50 nm pore polycarbonate filter. The vesicles are unilamellar, exhibit a narrow size dispersity and are characterized by a diameter of the order of the pore size. In **Supplementary Information S3**, it is also shown that the vesicle diameter varies linearly with the polycarbonate membrane porosity in the range 50 – 800 nm. These results show that extrusion is a technique that is able to reshape the Curosurf® multivesicular vesicles and provides means to assess the role of lipid assembly on the interaction with nanoparticles.

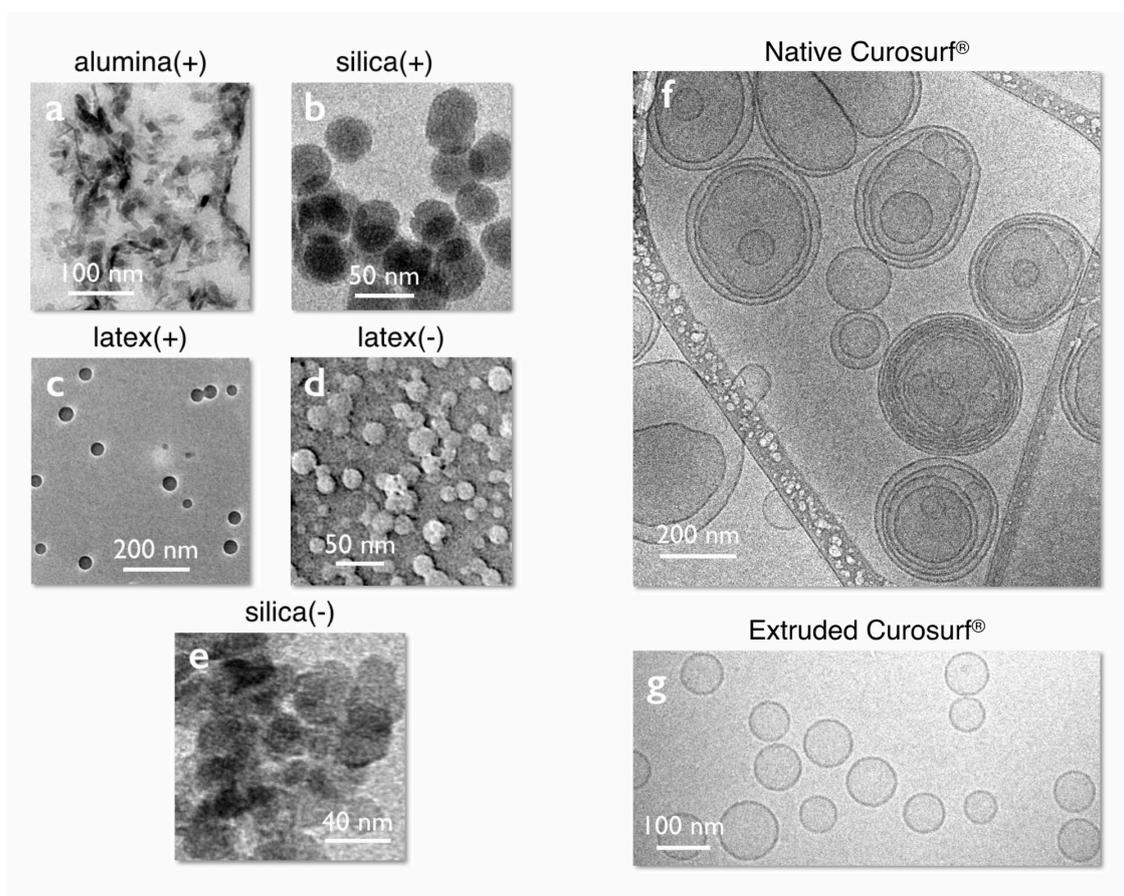

*Figure 1: a – e)* Transmission electron microscopy images of alumina, silica and latex particles used in this study. The particles are sorted according to their surface charge densities, which are $\sigma_{Alumina\,(+)}$ = +7.3e nm$^{-2}$, $\sigma_{Silica\,(+)}$ = +0.62e nm$^{-2}$, $\sigma_{Latex\,(+)}$ = +0.33e nm$^{-2}$, $\sigma_{Latex\,(-)}$ = -0.048e nm$^{-2}$ and $\sigma_{Silica\,(-)}$ = -0.31e nm$^{-2}$ (Table I). *f,g)* Cryo-TEM images of native and extruded (pore membrane 50 nm) Curosurf®. For the vesicles, the size of each of the enclosed structures has been included into the statistical analysis. For the native vesicles, the distribution is well described by a log-normal function of median 230 nm and dispersity 0.55.[37] For the extruded sample, the average vesicular size is 82 nm. Extra Curosurf® illustrations are provided in **S4 and S6**.

## III.2 – Job scattering plots

In 1928, Paul Job developed the method of continuous variation to determine the stoichiometry of binding (macro)molecular species in solutions, providing information about the equilibrium complexes.[44,45] We have adapted this technique to study nanoparticle-vesicle interactions using static and dynamic light scattering.[34,46,48,49,54,58] With this method, the total concentration of the mixed dispersion $c$ is held constant, whereas the fraction of each component is varied from $X = 0$ to $X = \infty$, where $X$ denotes the mixing ratio. Fig. 2 displays scattering intensity simulations for nanoparticles and vesicles *versus X* assuming 3 different interaction models, as detailed below. With these conventions, the particles are on the left hand-side ($X = 0$) and the vesicles are on the right ($X = \infty$).

We first consider the case where nanoparticles and vesicles do not mutually interact. The scattering intensity is thus the sum of the $X = 0$ and $X = \infty$ intensities, weighted by their actual concentrations $c_{NP}(X) = c/(1 + X)$ and $c_{Ves}(X) = cX/(1 + X)$ (continuous curve in grey).[46] In a second example, we consider that the particle-vesicle interaction is attractive, leading to the formation of mixed aggregates. The model is general and does not specify the interaction type. As the scattering varies linearly with the weight-averaged molecular weight of the scatterers, the presence of aggregates will lead to an enhanced scattering. This enhancement often takes the form of a peak centered on a given mixing ratio that reflects the aggregate stoichiometry.[46-48,58] Apart form the peak, the aggregates coexist with either single particles on the left-hand or with single vesicles on the right-hand sides. The simulated curve in orange is computed assuming the assembly of 200 nm unilamellar vesicles and 40 nm particles in a ratio 1:2 and that the association is electrostatic driven. Details are given in the caption and in **Supplementary Information S7**.

In the third model, it is assumed that the particle-vesicle interaction is still attractive, but there the lipid membrane spontaneously wraps around the particles and form a supported lipid bilayer. The spontaneous SLB formation has been suggested by recent experiments, in particular in the case of synthetic lipids and silica nanoparticle.[9,10,15,23,25,33] In the Job representation, the critical mixing ratio $X_C$ corresponds to the dispersion for which the particle and vesicle surface area concentrations are equal. In other words, at $X_C$ all particles are covered with the single lipid bilayer. $X_C$ is defined as $A_{NP}/A_{Ves}$ where $A_{NP}$ and $A_{Ves}$ are the nanoparticle and vesicle specific surface areas respectively. According to this model, SLB-coated particles coexist with single nanoparticles below $X_C$ and with single vesicles above. The SLB intensity was simulated for various configurations, varying the vesicle size and the scattering contrasts of the different species. The continuous green line in Fig. 2 was obtained from 40 nm particles and 200 nm vesicles using scattering contrasts in a ratio 1:0.8:0.5 for the particles, SLB coated particles and vesicles respectively. In this example and in every other inspected, the SLB scattering lies below that of the non-interacting species, indicating a peculiar behavior and a marked difference with the two other models. This remarkable result should allow us to distinguish which basic structures are formed upon mixing and attractive interaction, either aggregates or SLBs. In the following we define the interaction strength parameter $S_{Int}$ between particles and vesicles as the integral beneath the experimental scattering curve $I_S(X)$, relative to that of the non-interacting model:

$$S_{Int} = \frac{1}{I_S^{NP}(X=0)c} \int_0^\infty (I_S(X) - I_S^{non-Inter}(X))dX \quad (1)$$

To allow comparison between different nanoparticle systems, the integral is normalized with respect to the nanoparticle intensity and to $c$. It is thus expressed in L g$^{-1}$. According to Eq. 1 and to the results in Fig. 2, the aggregate and SLB formations are characterized by $S_{Int} > 0$ and $S_{Int} < 0$ respectively, whereas the non-interacting systems give $S_{Int} = 0$. In the next section light scattering was used to assess experimentally these different models.

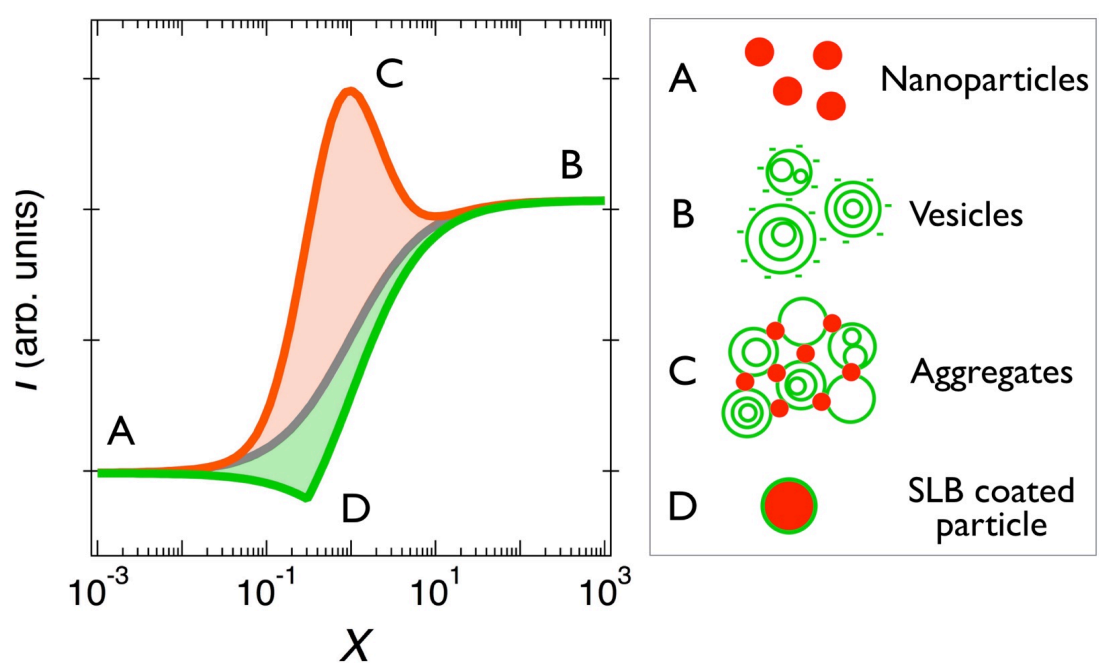

*Figure 2: Job scattering plots of mixed vesicle/particle dispersions according to various interaction models. X denotes the mixing ratio and $I_S$ the light scattering intensity. The continuous curves in grey, orange and green are predictions for the non-interacting species model, the aggregation of vesicles and particles and for the supported lipid bilayer formation respectively. Schematic representations of particles (A), multivesicular vesicles (B), aggregates (C) and SLB coated particles (D) are provided in the legend in the right-hand panel. The curve in orange is obtained assuming aggregation between 200 nm negatively charged unilamellar vesicles ($\sigma_{Vesicle}$ = -0.17e nm$^{-2}$) and 40 nm positively charges particles ($\sigma_{NP}$ = +0.62e nm$^{-2}$) in a ratio 1:2. The peak position is associated with the charge stoichiometry. The curve in green was obtained for 40 nm particles and 200 nm vesicles using scattering contrasts in a ratio 1:0.8:0.5 for the particles. In the Job scattering plots, the SLB formation is characterized by a decrease in the scattering intensity (details are provided in **Supplementary Information S7**).*

## III.3 – Interaction diagrams and strength parameters

In this part the interactions between nanoparticles and surfactant substitutes are investigated experimentally. The method of continuous variation is combined with the technique of static and dynamic light scattering to retrieve the scattering intensity $I_S(X)$ and hydrodynamic diameters $D_H(X)$ of the mixed dispersions. The data were first acquired at room temperature (T = 25 °C), *i.e* below the gel-to-fluid transition of Curosurf® occurring at $T_M$ = 29.5 °C (**Supplementary Information S8**). At this temperature, the phospholipids exhibit a long-range structural order, resulting in an increase of the membrane elasticity and bending energy.[59] It is anticipated that the wrapping of the bilayers around the particles will hence be less probable in these conditions. For the synthetic surfactant, we assume that the gel-to-fluid transition is closed to that of DPPC ($T_M$ = 41 °C) and that the temperature change between 25 °C and 37 °C should not affect the interaction with nanoparticles. The scattering data are shown in **Supplementary Information S9.1-5** for 14 pairs of nanoparticle/vesicle dispersions (**S9.1** for Alumina (+), **S9.2** for Silica (+), **S9.3** for Latex (+), **S9.4** for Latex (-) and **S9.5** for Silica (-)).

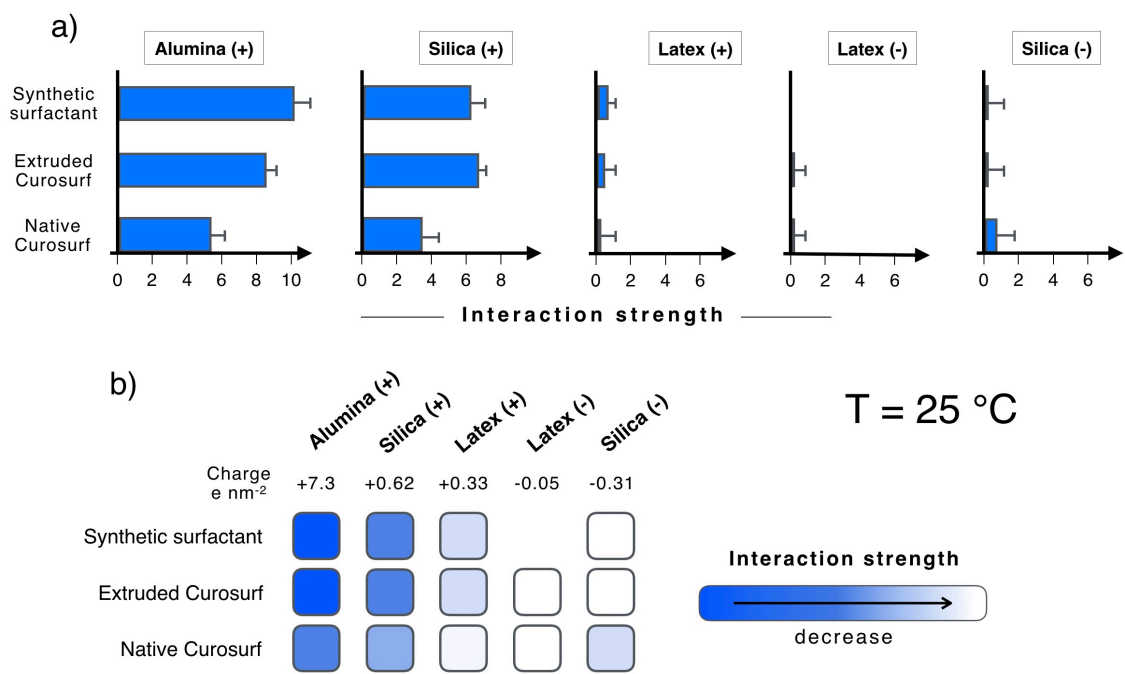

*Figure 3: a)* *Diagrams showing the interaction strength parameter $S_{Int}$ for the nanoparticle-surfactant dispersions at T = 25 °C. $S_{Int}$ is obtained from Eq. 1 and from the data in the Job scattering plots* **(Supplementary Information S9)**. *The interaction parameters are positive or zero, ruling out the spontaneous formation of supported lipid bilayers in the different cases tested.* **b)** *Schematic summary of the particle-vesicle interactions showing the correlation between interaction strength and positive charge densities. As surfactant membranes are negatively charged, it is concluded that the interaction is predominantly electrostatic.*

A close observation of the results reveals two types of behaviors. For particles such as Alumina (+) and Silica (+), and to a lesser extend for Latex (+), the intensities and diameters exhibit marked maxima, a feature that is reminiscent of the aggregate formation seen in Fig. 2. For the two others, Latex (-) and Silica (-), the scattering intensity varies continuously throughout the Job plot and is in good agreement with the non-interacting model. No instance of a negative peak indicative of SLB formation is identified. Fig. 3a displays a summary of the 14 data sets in terms of interaction strength parameter $S_{Int}$. It is found that for all conditions tested $S_{Int}$ is positive or zero and that for a given particle the interaction parameter is uniform amongst the three surfactants. With decreasing surface charge density, the average $S_{Int}$ are found to be +7.9, +5.4, +0.15, 0 and +0.11 L g$^{-1}$ for Alumina (+), Silica (+), Latex (+), Latex (-) and Silica (-) respectively. The results of Fig. 3a suggest that the role of the membrane proteins SP-B and SP-C is minor at this temperature and does not affect the overall interaction behavior. Fig 3b represents the light scattering data in terms of patches, from the dark blue for strong interacting systems to the light blue and white for non-interacting species. The figure evidences a clear correlation between the charge density and interaction strength parameter $S_{Int}$. This correlation, together with the formation of aggregates at a fixed mixing ratio suggests that pulmonary surfactant vesicles and nanoparticles interact predominantly via electrostatic complexation and opposite charges pairing. In case of same charge systems, with Latex (-) and Silica (-), interactions are not present.

The light scattering experiments were repeated at the physiological body temperature (T = 37 °C), *i.e.* above the gel-to-fluid transition of Curosurf®. Due to an increased flexibility of the membranes at this temperature, it is possible that the mixing with opposite charged surfaces induce the lipid fusion around the particles and eventually the formation of a SLB.[9,10,15,25,33] The scattering data are shown in **Supplementary Information S10.1-5** for 13 pairs of nanoparticle/vesicle dispersions, the sample numbering being similar to that of Fig. 3 (**S10.1** for Alumina (+), **S10.2** for Silica (+), **S10.3** for Latex (+), **S10.4** for Latex (-) and **S10.5** for Silica (-)). Fig. 4a displays the résumé of the intensity results in terms of the interaction strength parameter $S_{Int}$. It is found that the $S_{Int}$ parameter is positive or zero, indicating either the formation of aggregates (Alumina (+) and Silica (+)) or an absence of mutual interaction (Latex (-) and Silica (-)). For Latex (+) particles, the experimental scattering intensity in **Supplementary Information S10.3** is found to be lower than the prediction of the non-interacting model, leading to a negative $S_{Int}$ of -0.4 L g$^{-1}$. However this negative value is at the same time associated with the formation of aggregates, as noticed in the $D_H(X)$ data. The contradictory outcome found for Latex (+) may indicate the limit of the continuous variation method for strongly scattering particles. Apart form these later particles, we do not find evidence of spontaneous SLB formation with the native or extruded Curosurf® or with the synthetic surfactant. Compared to the 25 °C data, some differences are nevertheless noticeable. They concern Alumina (+) and Silica (+) for which the interaction parameter passes from $S_{Int}$ = 10 L g$^{-1}$ for synthetic surfactant to $S_{Int}$ = 2 L g$^{-1}$ with Curosurf®, either native or extruded. At a molecular level, these differences could be assigned to the phospholipid composition (Curosurf® possesses a wide variety of lipids, including DPPC, sphingomyelin and phosphatidylethanolamine[52]) and/or to the SP-B and SP-C proteins. More local investigations would be required to test this hypothesis. In conclusion, an extensive light scattering investigation has shown that SLB do not form spontaneously on nanoparticles of different kind. In contrast, we evidence that for opposite charge species, mixed vesicle/nanoparticle aggregates are formed, and remain stable over long periods of time (> weeks).

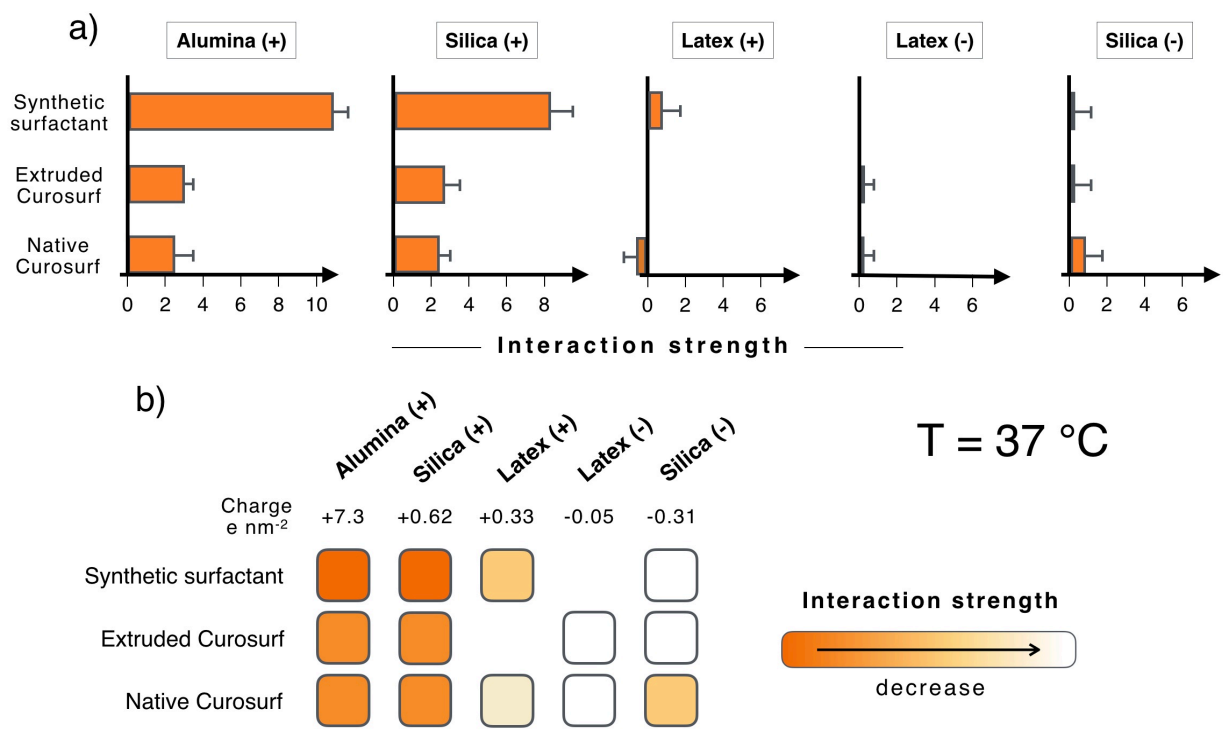

*Figure 4:* Same as Fig. 3 at T = 37 °C.

## III.4 – Structures of the vesicle/nanoparticle aggregates: the case of Silica (+) and Curosurf®

Here we address the issue of the internal structure of the aggregates found with positively charged nanoparticles and Curosurf®. **Fig. S11** shows phase-contrast optical microscopy images of mixed dispersions and on the right-hand side the corresponding size distributions. With Alumina (+) and Silica (+), large aggregates are visible and characterized by an average size of 6.5 and 3.7 µm respectively, i.e. larger than Curosurf® vesicles (1.8 µm), confirming hence directly the aggregation scenario. The data for Silica (-) are also presented for comparison and do not reveal the presence of micron-sized aggregates. Moreover, the average vesicular size agrees with that of the neat Curosurf® sample. The combination of light scattering and optical microscopy suggests that in case of strong association, the aggregates are broadly distributed in size, ranging from a few hundreds of nanometers to a few microns. To explore this extended length scale, two complementary techniques were used, fluorescence microscopy for objects larger than 500 nm and cryo-TEM for sub-micron structures.

In the sequel of the paper, we focus on Silica (+)/native Curosurf®, assuming it is representative of this class of aggregate forming materials. To prove co-localization, we take advantage of the optical properties of the aminated Silica (+) particles that fluoresce in the orange-red at 590 nm thanks to rhodamine molecules inserted in the silica network. In parallel, the Curosurf® vesicles are labeled with PKH67, a green fluorescent lipid used for tagging biological membranes and emitting at 502 nm.[53] Fig. 5A presents an extended view of a Silica (+)/Curosurf® dispersion observed under phase contrast (a), green (b) and red (c) illumination, illustrating aggregates from a few hundreds of nanometers to several microns (arrows). The merge image of Figs. 5Ab and 5Ac exhibits moreover an excellent superimposition of the green and red channels for most objects, indicating that the aggregates contain both fluorescent species (Fig. 5Ad). The inner aggregate structure can be seen further at higher magnification (Fig. 5B). In Figs. 5Ba-c, a cluster made of five large vesicles sticking to each other appears to fluoresce both in the green and red signals with a relative homogeneous distribution of the two colors. Figs. 5Bd-f display a 3 μm aggregate where green vesicles are visible and decorate the outer edge of the overall structure. These results, together with those of **Supplementary Information S12** ascertain that the aggregates are made of vesicles and particles and that both are intermixed at the micron scale. At this point, it can be assumed that the aggregates form via charge mediated association in a process where particles adhere to the vesicular membranes and develop links between them.[60]

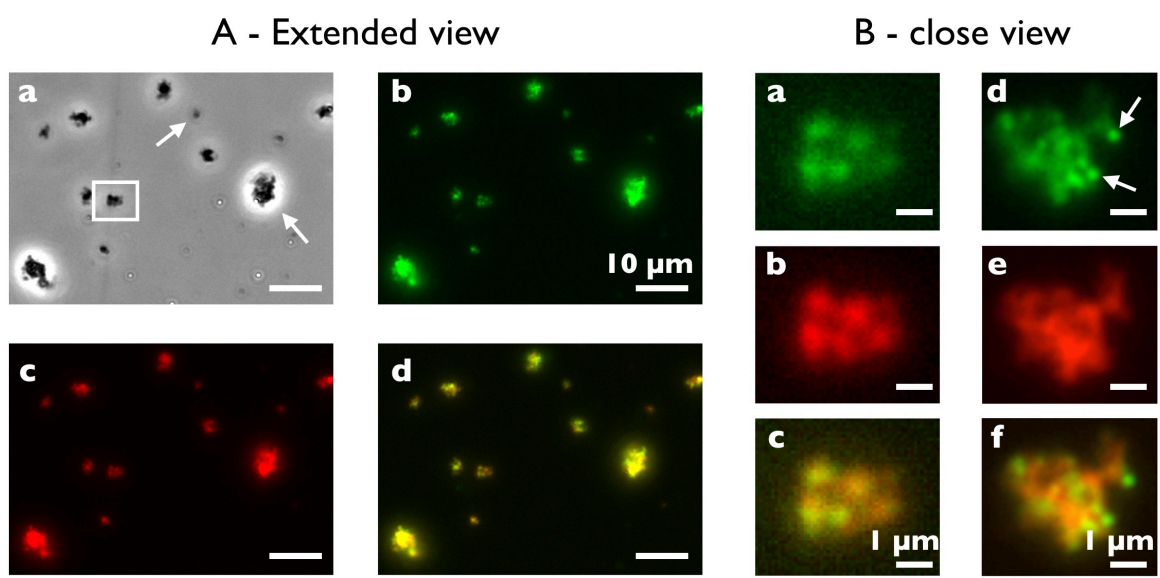

*Figure 5:* Extended (*A*) and close (*B*) illustration of aggregates made from 42 nm aminated silica nanoparticles and native Curosurf® observed by optical and fluorescence microscopy (magnification 60×). The experimental conditions are $c = 1$ g L$^{-1}$, $X = 2$ and $T = 37$ °C. The Silica (+) are synthesized to fluoresce in the orange-red at 590 nm and the vesicles are labeled with a green fluorescent lipid (PKH67) emitting at 502 nm. In *A*, a dispersion is observed under phase contrast (*a*), green (*b*) and red (*c*) illumination. In (*Aa*), the arrows point towards aggregates of various sizes. The white rectangle displays the 5-vesicle cluster later shown in (*Ba-c*). In *B*, examples of aggregates identified using green (*a,d*) and red (*b,e*) fluorescence. The merge signals are shown in (*Ad*) and in (*Bc, Bf*). Arrows in (*Bd*) indicate vesicles decorating the outer edge of the overall structure.

Fig. 6 displays cryo-TEM snapshots of particles and Curosurf® vesicles at the sub-micron length scale. For cryo-TEM experiments, particle and vesicle dispersions ($c$ = 1 g L$^{-1}$) were prepared at room temperature, then mixed and kept at 37 °C for 24 h. Following the grid preparation, samples were frozen in a fraction of a second. Different structures are observed, such as particle aggregates (Figs. 6a and 6c), particles sticking at the vesicular membrane (Figs. 6b and 6c) or internalized in vesicles (Figs. 6b, 6d and 6e). In this later case, the vesicles can be spherical or elongated, showing the membrane ability to deform and incorporate one to several particles. The elongated tubular structure in Fig. 6d resembles recent simulation predictions indicating that the cooperative wrapping of several nanoparticles in a tube membrane would be energetically more favorable than that of single nanoparticles.[16,22] The examples of Silica (+) adhering at the vesicle surface support the assumption that particles would act as physical links between membranes and be responsible of the aggregation. At higher magnification, supported lipid bilayers coating particles are also observed, however in limited numbers (Fig. 6e). It should be added that cryo-TEM also displays few particles and vesicles having not reacted with each other. In conclusion, cryo-TEM confirms the existence of strong attractive interaction between the aminated silica and vesicles. In addition to light scattering and optical microscopy which emphasize the formation of mixed aggregates, cryo-TEM discloses a broad variety of nanostructures that cannot be simply explained in terms of a unique type of interaction, here based on electrostatic charges, or formed following a single scheme. More experiments are needed to better understand the reorganization of the lipids at solid surfaces.

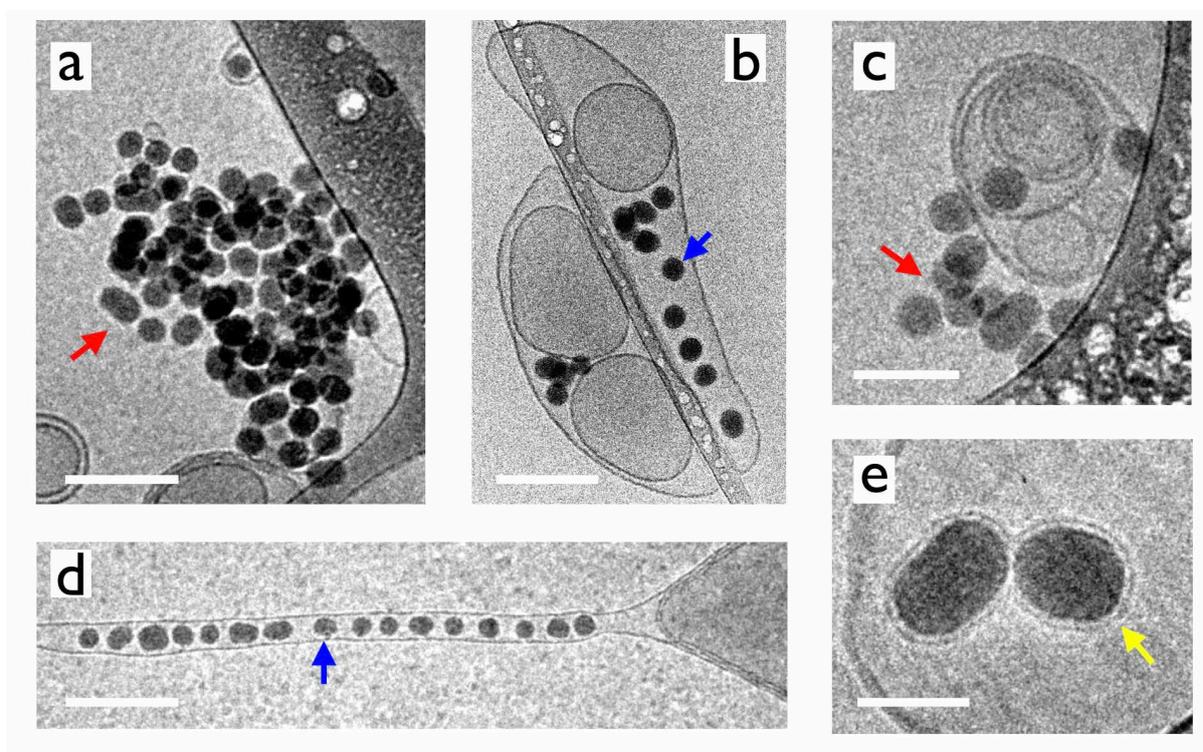

***Figure 6:*** *Cryo-TEM snapshots of 42 nm aminated silica nanoparticles and native Curosurf® prepared at room temperature and then mixed and kept at 37 °C for 24 h.* ***a, c)*** *particle aggregates, red arrows;* ***b,***

*c) particles sticking at vesicular membrane; **b, d, e)** particles internalized in vesicles, blue arrows; e) supported lipid bilayers, yellow arrow. Scales are 200 nm in **a**, **b** and **d**), 100 nm in **c**) and 40 nm in **e**).*

## IV – Conclusion

In this work we study the interaction of engineering nanoparticles with pulmonary surfactant substitutes. A mixture of synthetic phosphatidylcholine lipids and a drug administered to premature infants with respiratory distress syndrome, Curosurf® are formulated in controlled concentration, pH, temperature and ionic strength conditions. Curosurf® is studied in its native form and after extrusion through a porous membrane, leading to unilamellar vesicles. The uni- and/or multivesicular vesicles have size comprised between circa 100 nm to a few microns and are negatively charged, with zeta potentials varying from -23 mV to - 61 mV, depending on the preparation conditions. Pertaining to the nanomaterials, sub-100 nm aluminum oxide ($Al_2O_3$), silicon dioxide ($SiO_2$) or polymers (latex) particles are considered. The surface charge densities vary from +7.3$e$ nm$^{-2}$ for alumina particles to -0.31$e$ nm$^{-2}$ for negative silica. Here we focus on the phase behavior of mixed systems composed of nanoparticles, surfactant vesicles and a solvent over wide range of mixing conditions. Static and dynamic light scattering, optical fluorescence microscopy and cryo-transmission electron microscopy are used to evaluate the structures resulting from their mutual interaction. In relation with the method of continuous variation, we also provide theoretical calculations for the light scattering intensity in different strong interaction scenarios such as the formation of aggregates and of supported lipid bilayers. It is found that from the Job scattering representation, the SLB formation is associated with a decrease of the scattering intensity upon mixing, whereas the aggregate formation is associated with a scattering increase, allowing to distinguish between the two structures.

The central result of this work is the observation of mixed nanoparticle-vesicle aggregates resulting from electrostatic interaction. For both synthetic and exogenous surfactants, micron-sized hybrid aggregates are observed for positively charged alumina, silica and latex. The outcomes suggest that the SP-B and SP-C membrane proteins that are not present in the synthetic substitute have a minor role and do not affect the overall interaction behavior. In the cases of same charge systems *i.e.* with negatively charged latex and silica, direct or strong interactions are not observed. To assess the role of the membrane elasticity on the interaction, experiments were made below and above the gel-to-fluid transition of Curosurf®. To test the impact of uni- *versus* multivesicular vesicles on the interaction, Curosurf® was extruded using pore sizes between 100 and 800 nm and tested according to the same protocols. The results were similar to those obtained with native Curosurf®. Pertaining to the aggregate internal structure, fluorescence microscopy ascertains that the aggregates are made of vesicles and particles and that both are intermixed at the micron scale. Cryo-TEM confirms that the aggregates form *via* charge mediated association in a process where particles adhere to the vesicular membranes and develop links between them. In conclusion, we have seen that surfactant vesicles and particles interact strongly when they are of opposite charges and that the interaction is driven by electrostatics. Such aggregation processes may well occur in real-life situations for particles reaching the respiratory zone. Our approach also provides an illustration of potential molecular interactions that are at the origin of the protective role of surfactant against particles and pathogens. In particular, the aggregate formation is expected to slow down the nanoparticle diffusion in the hypophase and to mitigate their internalization in alveolar macrophages and pneumocytes. Although in biologically relevant conditions the amount of particles involved are lower than in the present conditions, this work highlights the fact that with particles acting as stickers between vesicles, inhaled nanomaterials could significantly modify the interfacial and bulk properties of the pulmonary surfactant and interfere with the lung physiology.

# Acknowledgments

We thank Armelle Baeza-Squiban, Victor Baldim, Mélody Merle, Mostafa Mokhtari, Evdokia Oikonomou, Chloé Puisney, Nicolas Tsapis for fruitful discussions. Annie Vacher and Marc Airiau from the Solvay Research & Innovation Centre Paris (Aubervilliers, France) are acknowledged for carrying out the cryo-TEM experiments. We also thank Stéphane Mornet from the Institut de Chimie de la Matière Condensée de Bordeaux (Université Bordeaux 1) for the synthesis of the aminated fluorescent silica nanoparticles. ANR (Agence Nationale de la Recherche) and CGI (Commissariat à l'Investissement d'Avenir) are gratefully acknowledged for their financial support of this work through Labex SEAM (Science and Engineering for Advanced Materials and devices) ANR 11 LABX 086, ANR 11 IDEX 05 02. This research was supported in part by the Agence Nationale de la Recherche under the contracts: ANR-13-BS08-0015 (PANORAMA), ANR-12-CHEX-0011 (PULMONANO) and ANR-15-CE18-0024-01 (ICONS) and by Solvay.

# TOC Figure

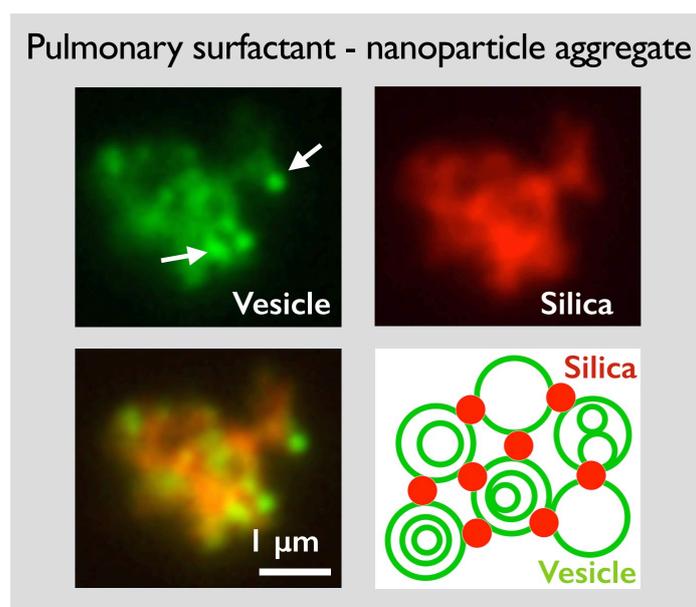

# References

1. R. D. Brook, S. Rajagopalan, C. A. Pope, J. R. Brook, A. Bhatnagar, A. V. Diez-Roux, F. Holguin, Y. L. Hong, R. V. Luepker, M. A. Mittleman, A. Peters, D. Siscovick, S. C. Smith, L. Whitsel and J. D. Kaufman, *Circulation*, 2010, **121**, 2331-2378.
2. T. Xia, Y. F. Zhu, L. N. Mu, Z. F. Zhang and S. J. Liu, *Nation. Sci. Rev.*, 2016, **3**, 416-429.
3. P. Bajaj, J. F. Harris, J. H. Huang, P. Nath and R. Iyer, *ACS Biomater. Sci. Eng*, 2016, **2**, 473-488.
4. E. Lopez-Rodriguez and J. Perez-Gil, *Biochim. Biophys. Acta, Biomembranes*, 2014, **1838**, 1568-1585.
5. R. H. Notter, *Lung surfactant: Basic science and clinical applications*, CRC Press, Boca Raton, FL, 2000.
6. J. Gil and E. R. Weibel, *Resp. Physiol.*, 1969, **8**, 13-36.
7. J. Goerke, *Biochim. Biophys. Acta, Mol. Basis Dis.*, 1998, **1408**, 79-89.
8. C. Dietrich, M. Angelova and B. Pouligny, *J. Phys. II*, 1997, **7**, 1651-1682.
9. O. Le Bihan, P. Bonnafous, L. Marak, T. Bickel, S. Trepout, S. Mornet, F. De Haas, H. Talbot, J. C. Taveau and O. Lambert, *J. Struct. Biol.*, 2009, **168**, 419-425.
10. J. W. Liu, *Langmuir*, 2016, **32**, 4393-4404.
11. A. E. Nel, L. Madler, D. Velegol, T. Xia, E. M. V. Hoek, P. Somasundaran, F. Klaessig, V. Castranova and M. Thompson, *Nat. Mater.*, 2009, **8**, 543-557.
12. H. Pera, T. M. Nolte, F. A. M. Leermakers and J. M. Kleijn, *Langmuir*, 2014, **30**, 14581-14590.
13. S. Savarala, S. Ahmed, M. A. Ilies and S. L. Wunder, *Langmuir*, 2010, **26**, 12081-12088.
14. F. Wang and J. W. Liu, *J. Am. Chem. Soc.*, 2015, **137**, 11736-11742.
15. F. Wang, X. H. Zhang, Y. B. Liu, Z. Y. Lin, B. W. Liu and J. W. Liu, *Angew. Chem.-Int. Edit.*, 2016, **55**, 12063-12067.
16. A. H. Bahrami, M. Raatz, J. Agudo-Canalejo, R. Michel, E. M. Curtis, C. K. Hall, M. Gradzielski, R. Lipowsky and T. R. Weikl, *Adv. Colloids Interface Sci.*, 2014, **208**, 214-224.
17. M. Deserno, *Phys. Rev. E Stat. Nonlin. Soft Matter Phys.*, 2004, **69**, 031903.
18. M. Deserno and W. M. Gelbart, *J. Phys. Chem. B*, 2002, **106**, 5543-5552.
19. R. Michel and M. Gradzielski, *Int. J. Mol. Sci.*, 2012, **13**, 11610-11642.
20. S. L. Zhang, H. J. Gao and G. Bao, *ACS Nano*, 2015, **9**, 8655-8671.
21. M. Raatz, R. Lipowsky and T. R. Weikl, *Soft Matter*, 2014, **10**, 3570-3577.
22. A. H. Bahrami, R. Lipowsky and T. R. Weikl, *Phys. Rev. Lett.*, 2012, **109**, 188102.
23. R. Michel, E. Kesselman, T. Plostica, D. Danino and M. Gradzielski, *Angew. Chem.-Int. Edit.*, 2014, **53**, 12441-12445.
24. E. Rascol, J. M. Devoisselle and J. Chopineau, *Nanoscale*, 2016, **8**, 4780-4798.
25. C. E. Ashley, E. C. Carnes, G. K. Phillips, D. Padilla, P. N. Durfee, P. A. Brown, T. N. Hanna, J. W. Liu, B. Phillips, M. B. Carter, N. J. Carroll, X. M. Jiang, D. R. Dunphy, C.

   L. Willman, D. N. Petsev, D. G. Evans, A. N. Parikh, B. Chackerian, W. Wharton, D. S. Peabody and C. J. Brinker, *Nat. Mater.*, 2011, **10**, 389-397.

26. L. De Backer, K. Braeckmans, J. Demeester, S. C. De Smedt and K. Raemdonck, *Nanomedicine*, 2013, **8**, 1625-1638.
27. L. De Backer, K. Braeckmans, M. C. A. Stuart, J. Demeester, S. C. De Smedt and K. Raemdonck, *J. Control. Release*, 2015, **206**, 177-186.
28. S. Dasgupta, T. Auth, N. S. Gov, T. J. Satchwell, E. Hanssen, E. S. Zuccala, D. T. Riglar, A. M. Toye, T. Betz, J. Baum and G. Gompper, *Biophys. J.*, 2014, **107**, 43-54.
29. G. Q. Hu, B. Jiao, X. H. Shi, R. P. Valle, Q. H. Fan and Y. Y. Zuo, *ACS Nano*, 2013, **7**, 10525-10533.
30. R. Ramanthan, K. Kamholz and A. M. Fujii, *J. Pulm. Resp. Med.*, 2013, **S13**.
31. X. B. Lin, T. T. Bai, Y. Y. Zuo and N. Gu, *Nanoscale*, 2014, **6**, 2759-2767.
32. P. N. Durfee, Y.-S. Lin, D. R. Dunphy, A. J. Muniz, K. S. Butler, K. R. Humphrey, A. J. Lokke, J. O. Agola, S. S. Chou, I. M. Chen, W. Wharton, J. L. Townson, C. L. Willman and C. J. Brinker, *ACS Nano*, 2016, **10**, 8325-8345.
33. S. Mornet, O. Lambert, E. Duguet and A. Brisson, *Nano Lett.*, 2005, **5**, 281-285.
34. F. Mousseau, R. Le Borgne, E. Seyrek and J.-F. Berret, *Langmuir*, 2015, **31**, 7346-7354.
35. A. K. Sachan, R. K. Harishchandra, C. Bantz, M. Maskos, R. Reichelt and H. J. Galla, *ACS Nano*, 2012, **6**, 1677-1687.
36. T. Curstedt, H. L. Halliday and C. P. Speer, *Neonatology*, 2015, **107**, 321-329.
37. F. Mousseau, C. Puisney, S. Mornet, R. Le Borgne, A. Vacher, M. Airiau, A. Baeza-Squiban and J. F. Berret, *Nanoscale*, 2017, **9**, 14967-14978.
38. C. Schleh, C. Muhlfeld, K. Pulskamp, A. Schmiedl, M. Nassimi, H. D. Lauenstein, A. Braun, N. Krug, V. J. Erpenbeck and J. M. Hohlfeld, *Respir. Res.*, 2009, **10**, 90.
39. S. Sweeney, B. F. Leo, S. Chen, N. Abraham-Thomas, A. J. Thorley, A. Gow, S. Schwander, J. F. J. Zhang, M. S. P. Shaffer, K. F. Chung, M. P. Ryan, A. E. Porter and T. D. Tetley, *Colloids Surf. B-Biointerfaces*, 2016, **145**, 167-175.
40. S. Vranic, I. Garcia-Verdugo, C. Darnis, J.-M. Sallenave, N. Boggetto, F. Marano, S. Boland and A. Baeza-Squiban, *Environ. Sci. Poll. Res.*, 2013, **20**, 2761-2770.
41. W. Wohlleben, M. D. Driessen, S. Raesch, U. F. Schaefer, C. Schulze, B. von Vacano, A. Vennemann, M. Wiemann, C. A. Ruge, H. Platsch, S. Mues, R. Ossig, J. M. Tomm, J. Schnekenburger, T. A. J. Kuhlbusch, A. Luch, C. M. Lehr and A. Haase, *Nanotoxicology*, 2016, **10**, 970-980.
42. M. Beck-Broichsitter, C. Ruppert, T. Schmehl, A. Guenther, T. Betz, U. Bakowsky, W. Seeger, T. Kissel and T. Gessler, *Nanomedicine*, 2011, **7**, 341-350.
43. D. Waisman, D. Danino, Z. Weintraub, J. Schmidt and Y. Talmon, *Clin. Physiol. Funct. Imaging*, 2007, **27**, 375-380.
44. P. Job, *Ann. Chim. France*, 1928, **9**, 113-203.
45. J. S. Renny, L. L. Tomasevich, E. H. Tallmadge and D. B. Collum, *Angew. Chem.-Int. Edit.*, 2013, **52**, 11998-12013.
46. J.-F. Berret, *Macromolecules*, 2007, **40**, 4260-4266.
47. J. Fresnais, C. Lavelle and J.-F. Berret, *J. Phys. Chem. C*, 2009, **113**, 16371-16379.

48. F. Mousseau, L. Vitorazi, L. Herrmann, S. Mornet and J. F. Berret, *J. Colloid Interface Sci.*, 2016, **475**, 36-45.
49. E. K. Oikonomou, F. Mousseau, N. Christov, G. Cristobal, A. Vacher, M. Airiau, C. Bourgaux, L. Heux and J. F. Berret, *J. Phys. Chem. B*, 2017, **121**, 2299-2307.
50. S. S. Raesch, S. Tenzer, W. Storck, A. Rurainski, D. Selzer, C. A. Ruge, J. Perez-Gil, U. F. Schaefer and C. M. Lehr, *ACS Nano*, 2015, **9**, 11872-11885.
51. N. Reinhardt, L. Adumeau, O. Lambert, S. Ravaine and S. Mornet, *J. Phys. Chem. B*, 2015, **119**, 6401-6411.
52. A. Braun, P. C. Stenger, H. E. Warriner, J. A. Zasadzinski, K. W. Lu and H. W. Taeusch, *Biophys. J.*, 2007, **93**, 123-139.
53. E. J. van der Vlist, E. N. M. Nolte-'t Hoen, W. Stoorvogel, G. J. A. Arkesteijn and M. H. M. Wauben, *Nat. Protoc.*, 2012, **7**, 1311-1326.
54. V. Torrisi, A. Graillot, L. Vitorazi, Q. Crouzet, G. Marletta, C. Loubat and J.-F. Berret, *Biomacromolecules*, 2014, **15**, 3171-3179.
55. J.-F. Berret, A. Sehgal, M. Morvan, O. Sandre, A. Vacher and M. Airiau, *J. Colloid Interface Sci.*, 2006, **303**, 315-318.
56. J. N. Israelachvili, *Intermolecular and Surfaces Forces*, Academic Press, New York, 1992.
57. A. Kumar, E. M. Bicer, A. B. Morgan, P. E. Pfeffer, M. Monopoli, K. A. Dawson, J. Eriksson, K. Edwards, S. Lynham, M. Arno, A. F. Behndig, A. Blomberg, G. Somers, D. Hassall, L. A. Dailey, B. Forbes and I. S. Mudway, *Nanomedicine*, 2016, **12**, 1033-1043.
58. L. Vitorazi, N. Ould-Moussa, S. Sekar, J. Fresnais, W. Loh, J. P. Chapel and J.-F. Berret, *Soft Matter*, 2014, **10**, 9496-9505.
59. Z. Yi, M. Nagao and D. P. Bossev, *J. Phys. Condens. Matter*, 2009, **21**.
60. S. Rose, A. Prevoteau, P. Elziere, D. Hourdet, A. Marcellan and L. Leibler, *Nature*, 2014, **505**, 382-385.